\documentclass[prl,amsmath,amssymb,aps,superscriptaddress,floatfix,reprint,notitlepage]{revtex4-1}
\usepackage{times}

\usepackage{amsmath, braket, amsfonts}
\usepackage{amssymb}
\usepackage{natbib}
\usepackage{sidecap}
\usepackage{bm, color, ulem}
\usepackage{xcolor}
\usepackage{lipsum}

\usepackage{graphicx}
\usepackage{dcolumn}
\usepackage{bm}

\begin{document}
\setcitestyle{super}
\preprint{APS/123-QED}




\title{Local probe of bulk and edge states in a fractional Chern insulator}

\author{Zhurun Ji}
\thanks{These two authors contributed equally}
\affiliation{Department of Physics, Stanford University, Stanford, 94305, CA, USA}
\affiliation{Department of Applied Physics, Stanford University, Stanford, 94305, CA, USA}
\affiliation{Stanford Institute for Materials and Energy Sciences, SLAC National Accelerator Laboratory, Menlo Park, 94205, CA, USA}

\author{Heonjoon Park}
\thanks{These two authors contributed equally}
\affiliation{Department of Physics, University of Washington, Seattle, Washington, 98195, CA, USA}
\author{Mark E. Barber}
\affiliation{Department of Physics, Stanford University, Stanford, 94305, CA, USA}
\affiliation{Department of Applied Physics, Stanford University, Stanford, 94305, CA, USA}
\affiliation{Stanford Institute for Materials and Energy Sciences, SLAC National Accelerator Laboratory, Menlo Park, 94205, CA, USA}

\author{Chaowei Hu}
\affiliation{Department of Physics, University of Washington, Seattle, Washington, 98195, CA, USA}

\author{Kenji Watanabe}
\affiliation{Research Center for Electronic and Optical Materials, National Institute for Materials Science,\\ 1-1 Namiki, Tsukuba, 305-0044Japan}
\affiliation{Research Center for Materials Nanoarchitectonics, National Institute for Materials Science,\\ 1-1 Namiki, Tsukuba, 305-0044, Japan}
\author{Takashi Taniguchi}
\affiliation{Research Center for Electronic and Optical Materials, National Institute for Materials Science,\\ 1-1 Namiki, Tsukuba, 305-0044Japan}
\affiliation{Research Center for Materials Nanoarchitectonics, National Institute for Materials Science, 1-1 Namiki, Tsukuba, 305-0044, Japan}

\author{Jiun-Haw Chu}
\affiliation{Department of Physics, University of Washington, Seattle, Washington, 98195, CA, USA}

\author{Xiaodong Xu}
\email{xuxd@uw.edu}
\affiliation{Department of Physics, University of Washington, Seattle, Washington, 98195, CA, USA}
\affiliation{Department of Materials Science and Engineering, University of Washington, Seattle, Washington, 98195, CA, USA}

\author{Zhi-xun Shen}
\email{zxshen@stanford.edu}
\affiliation{Department of Physics, Stanford University, Stanford, 94305, CA, USA}
\affiliation{Department of Applied Physics, Stanford University, Stanford, 94305, CA, USA}
\affiliation{Stanford Institute for Materials and Energy Sciences, SLAC National Accelerator Laboratory, Menlo Park, 94205, CA, USA}
\affiliation{Geballe Laboratory for Advanced Materials, Stanford University, Stanford, 94305, CA, USA}



\begin{abstract}
\textbf{Fractional quantum Hall effect (FQHE) is a prime example of topological quantum many-body phenomena, arising from the interplay between strong electron correlation, topological order, and time reversal symmetry breaking. Recently, a lattice analog of FQHE at zero magnetic field has been observed, confirming the existence of a zero-field fractional Chern insulator (FCI). Despite this, the bulk-edge correspondence — a hallmark of FCI featuring an insulating bulk with conductive edges — has not been directly observed. In fact, this correspondence has not been visualized in any system for fractional states due to experimental challenges. Here we report the imaging of FCI edge states in twisted MoTe$_2$ by employing a newly developed modality of microwave-impedance microscopy. By tuning the carrier density, we observe the system evolving between metallic and FCI states, the latter of which exhibits insulating bulk and conductive edges as expected from bulk-boundary correspondence. We also observe the evolution of edge states across the topological phase transition from an incompressible Chern insulator state to a metal and finally to a putative charge ordered insulating state as a function of interlayer electric field. The local measurement further reveals tantalizing prospects of neighboring  domains with different fractional orders. These findings pave the way for research into topologically protected 1D interfaces between various anyonic states at zero magnetic field, such as topological entanglement entropy, Halperin-Laughlin interfaces, and the creation of non-abelian anyons.}
\end{abstract}





\maketitle

\onecolumngrid



\section*{Introduction}

Fractional quantum Hall (FQH) states are distinguished by an incompressible two-dimensional bulk with edge channels that conduct fractional charge \cite{tsui1982two,laughlin1983anomalous,halperin1984statistics,arovas1984fractional,stern2008anyons}. While charge-sensing \cite{goldman1995resonant,martin2004localization,radu2008quasi}, shot noise \cite{de1998direct,bartolomei2020fractional}, and scanning gate microscopy \cite{pascher2014imaging} measurements have provided promising evidence of fractionalization, direct visualization of FQH edge states and the enclosed bulk remains elusive. Imaging these one-dimensional chiral edge modes is crucial to understand their physical properties, such as characteristic width and realizing Luttinger liquid physics \cite{chang2003chiral}. This could also significantly enhance our ability to manipulate edge state conduction for potential applications in anyonic braiding and topological quantum computing \cite{nayak2008non}. Microwave impedance microscopy (MIM) is a powerful local probe of material conductivity \cite{barber2022microwave} suitable for this purpose. MIM has been successfully employed to image edge states in several topological insulator systems, including quantum Hall \cite{lai2011imaging}, quantum spin Hall insulator \cite{shi2019imaging}, quantum anomalous Hall insulator \cite{allen2019visualization}, and graphene in the presence of quantizing magnetic field \cite{cui2016unconventional}. Despite these exciting progress, imaging fractional edge states with MIM is challenging due to the stringent experimental requirements for the formation of FQH states, such as high magnetic field and sub-kelvin temperature. 

The recent breakthrough in the realization of fractional quantum anomalous Hall effect (FQAHE) \cite{cai2023signatures,zeng2023thermodynamic,park2023observation,xu2023observation}, a zero field FQHE, provides an unique opportunity to image the fractional edge states. In addition to the absence of magnetic field, the FQAHE effect can survive to above 1 K. For instance, nearly quantized Hall resistance (\textit{R}$_{xy}$) with strongly suppressed longitudinal resistance (\textit{R}$_{xx}$) are observed for the $-2/3$ fractional Chern insulator (FCI) states at a temperature as high as 2~K in twisted MoTe$_2$ bilayer (t-MoTe$_2$) \cite{park2023observation}. These features position the zero-field FCI as a desirable platform for visualizing the fractional edge states. 

The experimental realization of FQAHE also accompanies several pressing questions. For instance, the transport measurements are mainly sensitive to edge state transport. Little is known about the bulk states' properties. The measured \textit{R}$_{xy}$ are not perfectly quantized while \textit{R}$_{xx}$ is suppressed but remain finite. These features are likely from the disorder of the material platform. In particular, the spatial variation of twist angle (i.e., moir{\'e} disorder) \cite{uri2020mapping}, will lead to domains and contribute to these observations. Phase transitions between Chern insulator states and topological trivial correlated insulator states have been inferred from transport measurements \cite{park2023observation,xu2023observation}. The nature of the phase transition and the properties of the state at the phase transition boundary remain to be understood. Further, the formation of fractional edge states are inferred from the transport with the observation of nearly quantized \textit{R}$_{xy}$ associated with suppressed \textit{R}$_{xx}$. A direct observation of the conducting edges with insulating bulk states at the fractional filled flat moir{\'e} Chern bands would serve as an important confirmation of the FCI states. As shown below, we successfully address these questions by a scanning MIM probe with sub-100~nm spatial resolution. This breakthrough is made possible by the recently developed exciton resonant-MIM and a novel gate design such that MIM imaging is possible through a top gate that is necessary to realize the desired physical phenomenon but otherwise prevents imaging study at the same time. 

\section{Results}\label{sec2}
\subsection{Experimental technique}

 We introduce the exciton resonant-microwave impedance microscopy (ER-MIM) technique to enable the sensing of dual-gated t-MoTe$_2$ samples. A prevalent challenge associated with the conventional MIM technique involves the measurement of devices with a top gate, which typically screens the probe, thereby hindering the direct assessment of the underlying sample. We successfully address this issue through both innovation in MIM technology and strategic sample design. The experimental configuration, depicted in Figure~1a, comprises a microwave transmission line, impedance-matched to a metallic scanning probe (tip) \cite{chu2020unveiling,barber2022microwave,cui2016quartz,huang2021correlated}, and a wavelength-tunable laser that is fiber-coupled into a Helium-3 cryostat. The device consists of a twisted MoTe$_2$ homobilayer with a twist angle of approximately 3.2$^{\circ}$, encapsulated within hBN layers and graphite as a bottom gate. Notably, rather than employing graphite or metal for the top gate, we have opted for a monolayer of tungsten disulfide (WS$_2$). Optical illumination of the entire sample near a photon energy resonant with the WS$_2$ A exciton ($\sim$2.2~eV) induces the optical gating and photoconductive effect (details in Ref. \cite{ji2023harnessing}), which allows the monolayer WS$_2$ to function as a tunable gate without screening microwaves from reaching the sample layer. All data are taken at zero magnetic field with a microwave frequency of 6.5~GHz. Our findings focus on the MIM-Im data, the imaginary component of the reflected microwave signal, which correlate with the local conductivity, or electronic compressibility \cite{wang2023probing} of the sample (see Methods for details).

\subsection{Local probe of bulk FCI states}

The inherent local probing capability of MIM allows us to explore the bulk state properties of fractional Chern insulators devoid of edge effects, a task that proves challenging through other means. Figure 1b shows the MIM-Im intensity as a function of carrier density (\textit{n}) and interlayer electric field (\textit{D}/$\epsilon$$_0$) that serves to tune the layer polarization of the moir{\'e} orbitals. The data is taken at a temperature of 1.5~K. The chosen color scheme highlights bulk insulating states in dark and conductive states in bright colors, revealing a rich landscape of electronic phases. Near \textit{D}/$\epsilon$$_0$ = 0, where the system assumes a honeycomb lattice, there are three prominent doping regimes with insulating behavior, centered at filling factors ($\nu$) $-$1, $-2/3$, and $-3/5$ respectively. Near $-1/2$ filling, the system is conductive, and eventually becomes insulating at low doping levels below a filling factor of $-1/3$. At large absolute values of \textit{D}/$\epsilon$$_0$, the charge carriers are fully polarized to a single layer forming a triangular lattice, and the system enters an insulating phase as seen from the dark line at commensurate fillings of $\nu$ = $-1$, and dark regions near $-2/3$ and $-1/2$. At intermediate electric fields, a bright conductive region separates the two insulating phases near zero and high \textit{D}/$\epsilon$$_0$. 

The MIM phase diagram of the bulk state presents a striking resemblance to the Chern insulator phase diagram derived from longitudinal resistance (\textit{R}$_{xx}$) measurements in transport studies. As detailed in Ref.~\cite{park2023observation}, a pronounced suppression of \textit{R}$_{xx}$ is observed in the quantum anomalous Hall (QAH) state near $\nu$ = $-$1 and the FCI state near $\nu$ = $-2/3$. This suppression is attributed to quantized edge state transport, indicating insulating bulk states --- a finding corroborated by our ER-MIM measurements. A finite but reduced \textit{R}$_{xx}$ near $\nu$ = $-1/2$ filling is linked to putative composite Fermi liquid states \cite{dong2023composite,goldman2023zero}, a behavior that aligns with the metallic response detected via MIM around this filling. The application of a perpendicular electric field at $\nu$ = $-1$, $-2/3$, and $-1/2$ induces a drastic increase in \textit{R}$_{xx}$, suggesting the emergence of either Mott or generalized Wigner crystal states \cite{regan2020mott,tang2020simulation,li2021imaging} due to full layer polarization. Our MIM measurement identifies an intermediate metallic phase amidst the electric field-induced phase transition from Chern insulator to charge-ordered states, evidence of a closure of the electronic gap during this topological phase transition. 

Leveraging the local probe capabilities of MIM, we park the tip at a pristine area of the sample and measure local conductivity. This effectively circumvents moir{\'e} disorder prevalent in electrical transport measurements, allowing more states to be resolved. Figure 1c plots MIM-Im as a function of filling factor under zero electric field at 1.5~K (orange) and 500~mK (blue), respectively. Distinct dips in conductivity are revealed at filling factors of $-1$, $-2/3$, and $-3/5$, indicating clearer FCI states compared to conventional transport measurements at the same temperature. Notably, at 500~mK, we observe a conductivity dip at $-4/7$ filling absent at 1.5~K, suggesting its energy gap is smaller than the other two FCI states. Furthermore, a subtle but appreciable conductivity dip at $\nu$ = $-1/2$ is observed at 500~mK. The weak dips near $-4/7$ , $-1/2$ and other features at even lower fillings are position dependent, and are only resolved in certain areas of the sample, which remains a future direction to explore with cleaner devices.  

\subsection{Imaging QAH edge state}

Scanning the metal tip enables us to explore the Chern insulator edge states. Figure 2a displays spatial maps of MIM-Im at selected filling factors, ranging from $-$1.28 to $-$0.73. The white arrow in Figure~2b indicates the parameter space for the panels in Figure~2a. The first panel of Figure~2a presents an AFM image of the sample, with the black dashed square highlighting the scan area. Below the grey dashed line lies the t-MoTe$_2$ sample region, which will be the focus of our discussion (See Extended Data Fig.1 for details). Starting at $\nu = -$1.28, the entire sample area exhibits a large MIM-Im signal, indicating it is in the metallic phase. As $|\nu|$ decreases, the signal within the bulk diminishes. At $\nu = -$1, the map becomes dark inside the bulk, while the edges light up. This is the exact behavior expected and observed from a QAH state, i.e., insulating in the bulk and conducting at the edge, consistent with bulk-boundary correspondence \cite{allen2019visualization}.  As the doping is further reduced, the t-MoTe$_2$ returns to a metallic state.   

We investigate the details of edge state evolution versus doping by extracting the spatial dependence of MIM-Im amplitude at selected filling factors. The results, depicted in Figure~2c, are taken along the dashed line shown in the second panel of Figure~2a.  The spatial linecuts reveal two notable features. First, the width of the spatial profile narrows as $\nu$ approaches $-$1, indicative of the formation of QAH edge states. However, the full width at half maximum of about 500 nm is much larger than the tip spatial resolution (see Extended Data Fig.2c for details). Second, the amplitude of the MIM-Im signal is higher in the QAH phase than the metallic phase. Both phenomena can be attributed to the emergence of gapless collective edge magnetoplasmon (EMP) modes from the 1D edge states circulating along the sample boundary, where the width of the edge MIM peak corresponds to the characteristic EMP length scale, comparable to the sample diameter \cite{wang2023probing}. 

 \subsection{Imaging fractional edge states}
 
Figure 3a shows MIM-Im spatial maps of the sample at carrier densities when the moir{\'e} Chern band is fractionally filled (along the blue arrow in Figure~2b). t-MoTe$_2$ becomes metallic when the system is slightly under-doped from the QAH state. This metallic state is evident by the strong MIM-Im signal over the entire sample area. Remarkably, as $\nu$ approaches the $-2/3$ filling, the map shows a dark bulk contrasted by bright edges, indicating that the corresponding $-2/3$ FCI state is insulating in the bulk and conductive at the edge. This pattern of edge conduction with an insulating bulk is also observed for the $-3/5$ FCI state. Figure 3b illustrates the evolution of the edge linecut signal as $\nu$ shifts from $-$0.71 (black curve) to the $-2/3$ FCI state (orange curve). Consistent with theoretical predictions \cite{lee1993quantum,wen1992theory}, this trend closely mirrors that of the QAH state. As $\nu$ approaches the $-2/3$ filling, the bulk becomes insulating and the edge conduction becomes pronounced.

We also compare the edge state signals between the FCI and QAH states. As depicted in Figure~3c, the black, purple, and yellow curves represent the MIM-Im measurements at $\nu = -1$, $-2/3$, and $-3/5$, under the same conditions (see Extended Data Fig. 3 for spatial maps). The former and the latter two correspond to QAH and FCI states, respectively. The data reveal that although the width of the FCI edge peak is similar to that of the integer state, the $-2/3$ and $-3/5$ FCI states exhibit higher MIM-Im signals at the edge compared to the QAH state. This observation does not align with the quantization of $\sigma_{xy}$ and cannot be solely explained by differences in bulk dissipation, as evidenced by the similar edge signals but distinct bulk signals in the $-2/3$ and $-3/5$ FCI states. One plausible explanation could be the differing velocities, hence resonance frequencies of magnetoplasmons between the FCI and QAH states, due to their distinct charges (e.g., two copies of $e^*=e/3$ versus $e$). This understanding is in agreement with prior microwave resonance measurements in a quantum well-based fractional quantum Hall systems \cite{wassermeier1990edge}. Additional complexities, such as edge reconstruction \cite{wen1992theory}, may also play a role, underscoring the need for further theoretical investigation.

The scanning capability of MIM exposes spatial inhomogeneities in the sample, evident from the spatial variation of the MIM-Im signal. This is likely due to moir{\'e} disorder and non-uniform gating potentials from remote defects in the WS$_2$ layer. Taking advantage of the high spatial resolution of the metal tip, the ER-MIM serves as a potent local transport tool, unveiling physics that is challenging to detect in global measurements. We demonstrate this capability by observing adjacent domains of different FCI states, with the interface defined by a narrow channel in the top gate. Figure 3d presents the MIM-Im signal versus gate voltage from two regions, marked by squares in the last panel of Fig.~3a. Due to the differences in gating efficiency or Schottky barrier height in these two areas (see Methods and Extended Data Fig. 1), the $-3/5$ FCI state in regime 1 coexists with the $-2/3$ state in regime 2. Future designs, such as split top gates, could enable independent control over these FCI states, facilitating the study of topologically protected 1D interfaces formed between different anyonic orders at zero magnetic field. These possibilities include the edge state scattering/coupling between various FCI states (e.g. between Halperin and Laughlin states), topological entanglement entropy of gaped 1D states, creation of new anyonic states, and future anyonic braiding operations \cite{nakamura2020direct,nakamura2023fabry,santos2018symmetry,crepel2019model,crepel2019microscopic}. 

\subsection{Edge state evolution across topological phase transitions}

Finally, we explore how the edge states evolve across topological phase transitions induced by an out-of-plane electric field. Figure 4a shows the MIM-Im signal versus electric field of the $\nu=-3/5$ (red, top), $-2/3$ (orange, middle) FCI states, and $-1$ (purple, bottom) QAH state. The MIM signal is suppressed at both low and high $|D/\epsilon_0|$, but is pronounced in the intermediate regime. As previously discussed, this suggests a continuous topological phase transition from the Chern insulator state to a metallic state, and eventually to a correlated insulating state as $|D/\epsilon_0|$ increases from zero to a large value. Figure 4b depicts the MIM-Im image at selected electric fields for the $-2/3$ FCI state. At high electric fields, the system is in a correlated insulating state, resulting in a dim overall MIM-Im image. At the phase transition boundary, the image brightens, reflecting the system entering a metallic state as the gap closes. Near \textit{D}/$\epsilon$$_0$ = 0, t-MoTe$_2$ is in the FCI state with a dark signal in the bulk and bright edges. The striking contrast of the presence and absence of the edges at FCI and the trivial insulating state confirms that the observed edge state for the FCI is not due to the trivial charge accumulation at the sample edges, but rather a manifestation of bulk-boundary correspondence. The observation of the conductive channel at the domain boundary between the two different FCI states in Figure~3 also supports our understanding. Meanwhile, the distinction of non-trivial edges is evident for the $-1$ QAH state as expected (Extended Data Fig.5). On the other hand, as shown in the Extended Data Fig. 6, the $-1/2$ state at 1.5~K experiences a transition from a compressible state at small $|D/\epsilon_0|$ to an incompressible state at large $|D/\epsilon_0|$.

\section{Summary} 
Using ER-MIM as a local probe combined with a monolayer WS$_2$ top gate structure, we realize the first MIM measurements of \textit{dual} gated devices at \textit{zero} magnetic field. This experimental breakthrough enables the visualization of both QAH and FCI states in t-MoTe$_2$, characterized by an insulating bulk and metallic edge, providing strong confirmation for the previous discoveries made through transport measurements \cite{park2023observation,xu2023observation}. This conclusion is further substantiated by the clear distinction between the FCI states and trivial charge-ordered insulators via a topological phase transition tuned by electrical field. The imaging capability allows for comparative studies of the edge states between fractional and integer states, as well as among the various fractional states. The local nature of the probe also reveals rich phenomena that is difficult to discern in transport measurements due to disorder, including the proximity of nearby regions of distinct fractional orders and the more subtle fractional states. These observations point to exciting future directions based on the FQAHE using highly controllable material platforms.

\section{Methods}
\subsection{Sample Fabrication}

Single crystals of 2H-MoTe$_2$ were grown by Te self-flux. H$_2$ annealed Mo powder and Te lumps were first mixed at molar ratio near 1:200 and sealed in a quartz tube in vacuum. The tube was then placed in a muffle box furnace, which was heated up to 750\textdegree C, and slowly cooled to 500\textdegree C at a rate of 1 K per hour. Finally, the crystals were separated from the flux, and annealed in vacuum at 500\textdegree C for a day to further remove the residual Te flux. 

Full details of the sample fabrication procedure can be found in Extended Data Fig.1. Hexagonal boron nitride (hBN) and graphite flakes were exfoliated on Si/SiO$_2$ substrates and characterized using contrast-enhanced optical microscopy and atomic force microscopy (AFM). First, the bottom gate structure was fabricated using a standard poly-(bisphenol A) carbonate-based dry transfer process. An hBN bottom gate dielectric was picked up, successively followed by a graphite bottom gate electrode, and melted down on an intrinsic silicon substrate. The polycarbonate was washed off using chloroform and dichloromethane. The bottom gate was prepatterned using standard e-beam lithography, and Ti/Pt (2/5 nm) electrodes and Cr/Au (5/60 nm) conductive pads were deposited with e-beam evaporation. Next, a monolayer MoTe$_2$ flake was exfoliated onto a silicon substrate inside a glovebox with O$_2$ and H$_2$O levels less than 0.1 ppm. The monolayer flake was cut in half using an AFM tip before the transfer to minimize strain. The twisted MoTe$_2$ heterostructure was created by picking up part of the monolayer with a hBN top gate dielectric, rotating the transfer stage by a desired angle, and picking up the remaining flake, before melting down onto the prepatterned bottom gate. Subsequently, an additional lithography and metal evaporation (Ti/Pt and Cr/Au) was performed to create electrodes and pads for the top gate. Finally, a monolayer WS$_2$ top gate was picked up and melted down on the stack to complete the dual-gated heterostructure. Contact-mode AFM was used to clean the surface in between every step of the process to ensure the absence of polymer residue and bubbles. During the final step of cleaning, the AFM tip created a narrow channel within the monolayer WS$_2$ top gate that allows the formation of different Chern insulator domains. 

\subsection{Estimation of filling factor based on doping density}

The carrier density $n$ and electric field $D$ on the sample were calculated from the top gate voltage V$_{\text{tg}}$ and bottom-gate voltage V$_{\text{bg}}$ using the equations $n=(V_{\text{tg}}C_{\text{tg}}+V_{\text{bg}}C_{\text{bg}})/e - n_{\text{offset}}$ and $D/\epsilon_0=(V_{\text{tg}}C_{\text{tg}}-V_{\text{bg}}C_{\text{bg}})/2\epsilon_0- D_{\text{offset}}/\epsilon_0$, where $e$ is the electron charge, $\epsilon_0$ is the vacuum permittivity, and $C_{\text{tg}}$ and $C_{\text{bg}}$ are the top and bottom-gate capacitances, respectively. The capacitances were determined from the gate thickness measured by atomic force microscopy. $D_{\text{offset}} \approx 0$ was inferred from the dual-gate map. The sequence of fractional fillings with prominent features in the dual-gate MIM-Im map (Fig. 1b) was used to identify the moiré filling factor. 

\subsection{Microwave Impedance Microscopy measurements}
MIM measurements were performed in a 3-He cryostat with a 12 T superconducting magnet and homebuilt scanning setup. The MIM probe, an etched tungsten wire, was attached to a quartz tuning fork for topographic sensing and scans were taken with the tip held approximately 40 nm above the sample’s surface. MIM measures changes in admittance between the tip and sample at GHz frequencies which can be related to changes in local conductivity and permittivity in the sample (see Extended Data Fig. 2b for details). The input power is $\sim$0.2$\mu$W. The measurements reported here were carried out at 6.5 GHz at zero magnetic field. The optical coupling for ER-MIM was realized with a supercontinuum laser (NKT photonics, FIU-20). The wavelength was selected with a monochromator (Princeton instrument, Acton SP 2300), with a linewidth of around 0.5 nm. The laser power at the end of the 100 $\mu$m diameter multimode fiber was $\sim$0.1--1~mW throughout the measurements, with the spot radius at the end of the tip being around 0.8 mm.

\subsection{Finite-element simulation}
Finite-element analysis (FEA) of the MIM response to 2D sheet conductance is shown in Fig.~\ref{Extended Data fig:2}. The tip radius is chosen to be 50 nm, and the SiO$_2$ layer thickness 300 nm with $\kappa_{\text{SiO}_2}$ = 3.9. The MIM-Im response curve as a function of the sheet conductance of $\sigma_{t-\text{MoTe}_2}$ is simulated. Both simulations with and without the gates were conducted, and they differ by a scaling factor, so the simulation result is shown in arbitrary units. 

\section{Acknowledgements}
We thank Dunghai Lee, Steve Kivelson and Taige Wang for stimulating discussions. This work at the Stanford Institute for Materials and Energy Sciences (SIMES) is supported by the QSQM, an Energy Frontier Research Center funded by the U.S. Department of Energy (DOE), Office of Science, Basic Energy Sciences (BES), under Award \#DE-SC0021238. Previous development of MIM technique at Stanford University was funded in part by the Gordon and Betty Moore Foundation’s EPiQS Initiative through Grant GBMF4546 to ZX.S. Z. J. acknowledges support from the Stanford Science fellowship and the Urbanek-Chodorow fellowship. The work at University of Washington is mainly supported by DoE BES under award DE-SC0018171. Bulk MoTe$_2$ crystal growth is supported by the Center on Programmable Quantum Materials, an Energy Frontier Research Center funded by DOE BES under award DE-SC0019443. Device fabrication used the facilities and instrumentation supported by NSF MRSEC DMR-230879.  K.W. and T.T. acknowledge support from the Elemental Strategy Initiative conducted by the MEXT, Japan (Grant Number JPMXP0112101001) and JSPS KAKENHI (Grant Numbers 19H05790, 20H00354 and 21H05233). 

\section{Author contributions}
ZX.S. and XD.X. initiated the collaboration. Z.J., H.P., XD.X. and ZX.S. conceived the design of the experiment. Z.J. and M.B. carried out the ER-MIM measurements. H.P. fabricated the device. C.H. and JH.C. synthesized the bulk MoTe$_2$ crystals. K.W. and T.T. provided bulk hBN crystals. Z.J., H.P., XD.X. and ZX.S. wrote the manuscript with inputs from all authors.

\section{Competing Interests}
ZX.S. is a co-founder of PrimeNano Inc., which licensed the MIM technology from Stanford University. The other authors declare no competing interests. 

\section{Data Availability}
The datasets generated during and/or analyzed during this study are available
from the corresponding author upon reasonable request. 

\normalem
\let\oldaddcontentsline\addcontentsline
\renewcommand{\addcontentsline}[3]{}
\bibliographystyle{naturemag}

\setcounter{figure}{0}

\renewcommand{\figurename}{\textbf{Figure}}
\renewcommand{\thefigure}{\arabic{figure}}
\renewcommand{\thefigure}{{\bfseries \arabic{figure}}}
\makeatletter
\renewcommand{\@caption@fignum@sep}{{\bfseries.} }
\makeatother
\newpage
 \begin{figure}[h!]
 \centering
  \includegraphics[width=1\textwidth,trim=0 0 0 0, clip]{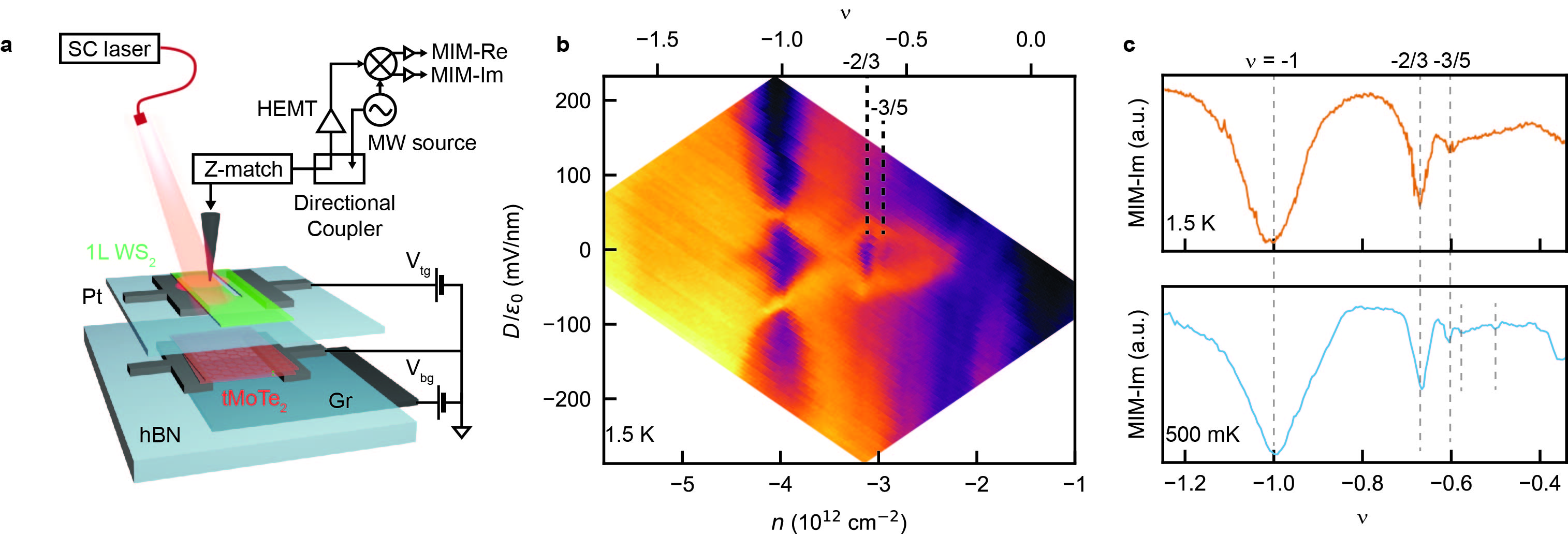}
 \caption{\textbf{Local probe of bulk fractional Chern insulator (FCI) states.} \textbf{a}. Schematic of the exciton-resonant microwave impedance microscopy setup, and the t-MoTe$_2$ device. The dual gated device has a monolayer WS$_2$ top gate (V$_{\text{tg}}$) and a graphite bottom gate (V$_{\text{bg}}$). \textbf{b}. The MIM-Im signal as a function of electric field ($D/\epsilon_0$) and carrier density ($n$) at 1.5~K. The filling factor ($\nu$) is shown on the top axis (See text and methods for details.) \textbf{c}. MIM-Im signal versus filling factor $\nu$ at $T = 1.5$~K (top, orange curve), and $T = 500$~mK (bottom, blue curve). The suppression of the signal at several filling factors signifies the bulk insulating properties of Jain sequence FCI states.}
  \label{fig:1}
 \end{figure}

\newpage
 \begin{figure}[h!]
 \centering
  \includegraphics[width=1\textwidth,trim=0 0 0 0cm, clip]{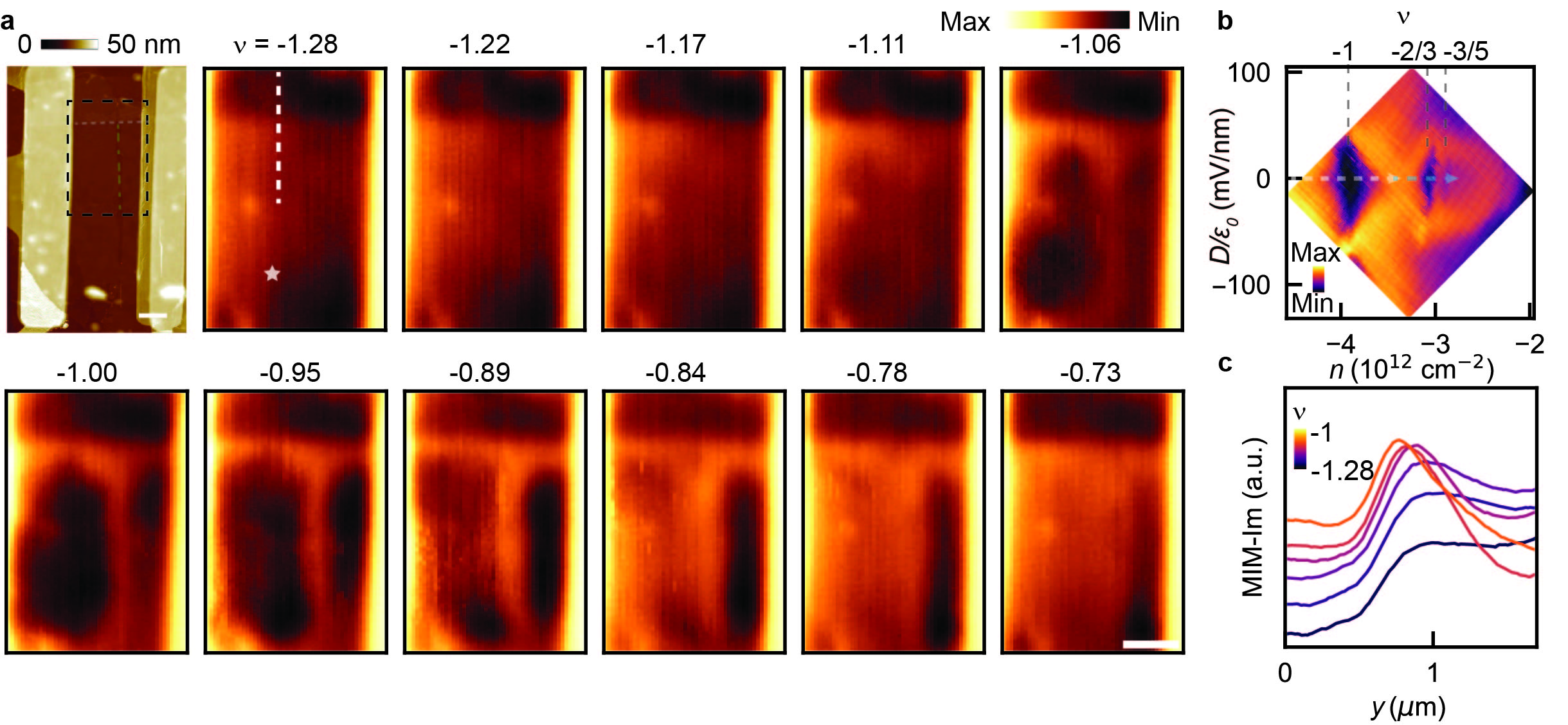}
 \caption{\textbf{Imaging quantum anomalous Hall edge states.}  \textbf{a}, First panel: an AFM image of the sample, showing the edges and interfaces (gray and green dashed lines, respectively) and the scanning region (black rectangle). Scale bar: 1~$\mu$m. Other panels: Real-space maps of the imaginary microwave response (MIM-Im) at different fillings, showing bulk-edge signal contrast being developed near filling $\nu=-1$. Scale bar: 1~$\mu$m. \textbf{b}, The MIM-Im signal as a function of electric field ($D/\epsilon_0$) and carrier density ($n$) at 1.5~K, measured at the spot denoted by the white star in \textbf{a}. The filling factor ($\nu$) is shown on the top axis. The white arrow denotes for the electric field-carrier density conditions for obtaining images in Fig. 2a, and the blue arrow denotes for that of Fig. 3a. \textbf{c}. The linecuts averaged over 0.7~$\mu$m width around the white dashed line in Fig.~2\textbf{a} of MIM-Im signals (with vertical offsets) at different filling factors traversing the physical edge.}
  \label{fig:2}
 \end{figure}

 \newpage
 \begin{figure}[h!]
 \centering
  \includegraphics[width=1\textwidth,trim=0 0cm 0 0cm, clip]{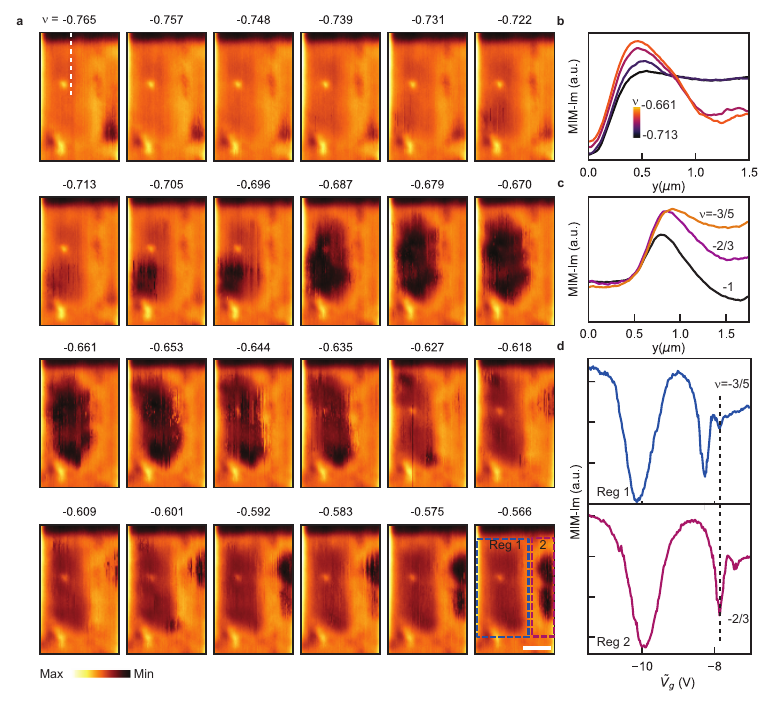}
 \caption{\textbf{Imaging fractional edge states.}  \textbf{a}, Real-space maps of the imaginary microwave response at different fillings, showing bulk-edge signal contrast being developed near filling $\nu = -2/3$ and $\nu = -3/5$. Scale bar: 1~$\mu$m. \textbf{b}, The linecuts of MIM-Im signals averaged over 0.3~$\mu$m width around the white dashed line in Fig.~3\textbf{a} (in Region 1), at a series of different filling factors between -0.661 and -0.713. \textbf{c}, The linecuts of MIM-Im signals traversing the edge at $\nu =$ -2/3, -3/5  and -1, averaged over 0.5~$\mu$m width. Original data in Extended Data Figure \ref{Extended Data fig:3}. \textbf{d}, MIM-Im signal versus $\Tilde{V}_{\text{g}} = V_{\text{bg}}+(C_{\text{tg}}/C_{\text{bg}})V_{\text{tg}}$ at two representative locations in the two regions of the sample at $T$ = 1.5~K, showing the coexistence of the two FCI states at $\nu =-3/5$ and $\nu =-2/3$.}
  \label{fig:3}
 \end{figure}

 \newpage
 \begin{figure}[h!]
 \centering
  \includegraphics[width=1\textwidth,trim=0 0cm 0cm 0cm, clip]{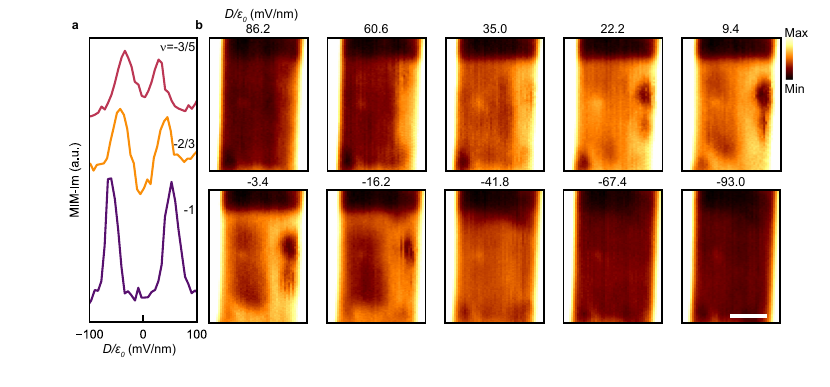}
 \caption{\textbf{Edge state evolution across electric field induced topological phase transition.}  \textbf{a}, MIM-Im signal as a function of $D/\epsilon_0$ at $T$ = 1.5 K. Red, yellow, and purple curves plotted with vertical offsets correspond to the filling factor $-$3/5, $-$2/3, and $-$1 Chern insulator state. \textbf{b}, Real-space maps of the MIM-Im at different displacement fields at $\nu = -2/3$ filling factor, showing bulk-edge signal contrast being developed during the formation of the FCI state near $D = 0$. Scale bar: 1$\mu$m. }
  \label{fig:4}
 \end{figure}

\setcounter{figure}{0}

\renewcommand{\figurename}{\textbf{Extended Data Figure}}
\renewcommand{\thefigure}{\textbf{\arabic{figure}}}
\makeatletter
\renewcommand{\@caption@fignum@sep}{{\bfseries.} }
\makeatother
 \newpage
 \begin{figure}[h!]
 \centering
  \includegraphics[width=1\textwidth,trim=0 0cm 0 0cm, clip]{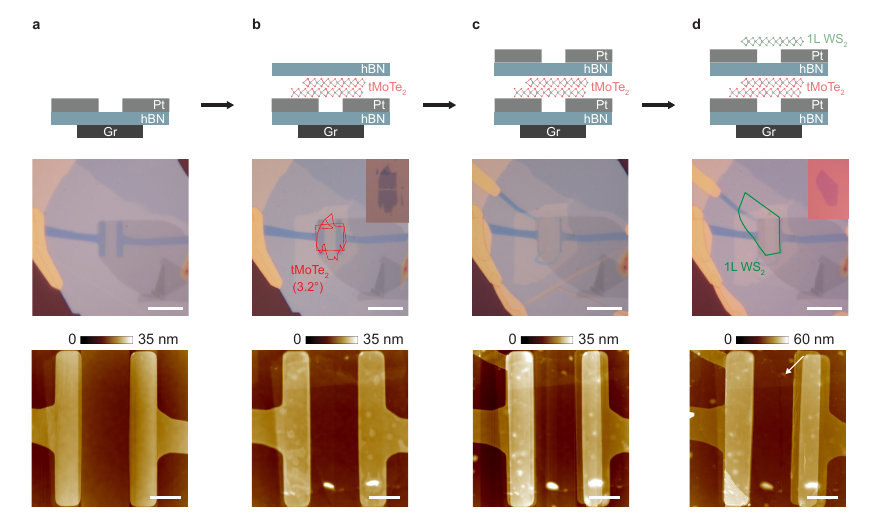}
\caption{\textbf{Device fabrication procedure} Each row indicates the device schematic, optical image, and AFM image starting from the top. \textbf{a}, hBN/graphite bottom gate with prepatterned platinum (Pt) electrodes. \textbf{b}, An hBN/t-MoTe$_2$ stack is put down on the bottom gate. Inset: optical image of 1L MoTe$_2$ flake that is cut in half using an AFM tip. \textbf{c}, Pt electrodes are patterned on top. \textbf{d}, A 1L WS$_2$ flake is put down to serve as a top gate. Inset: optical image of 1L WS$_2$ flake. The white arrow indicates the tear within the 1L WS$_2$ top gate that defines different Chern insulator domains. Scale bar in optical images and AFM images are 10 $\mu$m and 2 $\mu$m, respectively. 
 }
 \label{Extended Data fig:1}
 \end{figure}

\begin{figure}[h!]
 \centering
\includegraphics[width=0.8\textwidth,trim=0 0cm 0 0cm, clip]{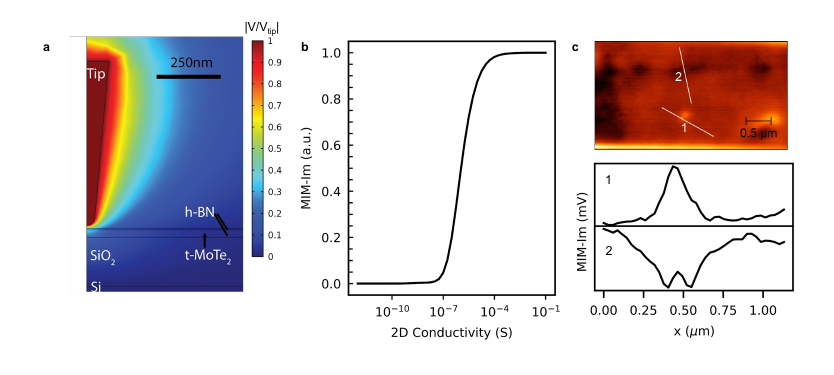}
\caption{\textbf{MIM measurement details} \textbf{a}, Tip and sample device geometry, and finite element calculations of the quasi-static potential distribution ($V$). The colorplot is $|V/V_{\text{tip}}|$, with $\sigma=10^{-12}$ S$\cdot$sq. \textbf{b}, The finite element simulation result of MIM-Im response curve as a function of the 2D conductivity of twisted MoTe$_2$ sample. \textbf{c}, Two MIM-Im line profiles showing that the measurement can resolve features with full width half maximum of $\sim$100 nm. }
 \label{Extended Data fig:2}
 \end{figure}

 \newpage
\begin{figure}[h!]
 \centering
\includegraphics[width=0.5\textwidth,trim=0 0cm 0 0cm, clip]{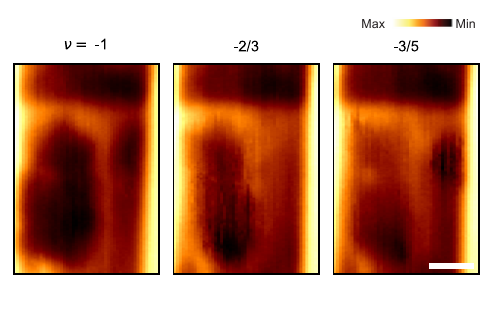}
\caption{\textbf{Real space maps of MIM-Im response for edge signal comparison.} Real-space maps of the MIM-Im response at different fillings measured at $\nu = -1$, $-2/3$ and $-3/5$ for edge signal comparison. Scale bar: 1$\mu$m. }

 \label{Extended Data fig:3}
 \end{figure}

\begin{figure}[h!]
 \centering
\includegraphics[width=1\textwidth,trim=0 0cm 0 0cm, clip]{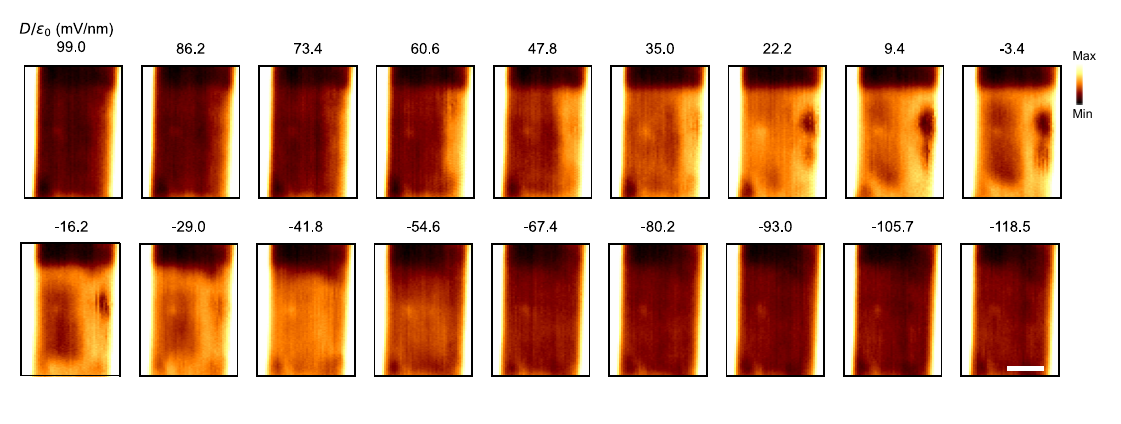}
\caption{\textbf{Real space maps of MIM-Im response at different displacement fields at $\bm{\nu = -2/3}$ filling.} Real-space maps of the MIM-Im at different displacement fields at $\nu = -2/3$ filling factor, showing the topological phase transition measured at $T$ = 1.5~K. Scale bar: 1$\mu$m.  }
 \label{Extended Data fig:4}
 \end{figure}

\begin{figure}[h!]
 \centering
\includegraphics[width=0.8\textwidth,trim=0 0cm 0 0cm, clip]{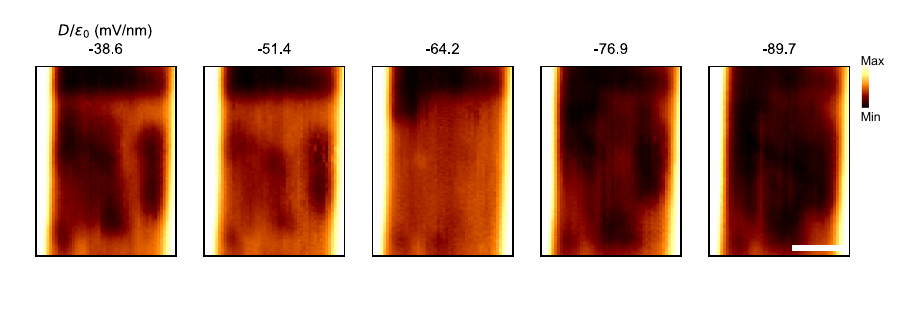}
\caption{\textbf{Real space maps of MIM-Im response at different displacement fields at $\bm{\nu = -1}$ filling.}  Real-space maps of the MIM-Im at different displacement fields at $\nu = -1$ filling factor, presenting the transition from QAH state at small displacement field, to metallic, then to trivial insulating state at larger displacement fields measured at $T$ = 1.5~K. The phase coexistence at the phase boundary is observed as the spatially inhomogeneous MIM-Im signal at certain displacement fields. Scale bar: 1$\mu$m.  }
 \label{Extended Data fig:5}
 \end{figure}
 
 \begin{figure}[h!]
 \centering
\includegraphics[width=1\textwidth,trim=0 0cm 0 0cm, clip]{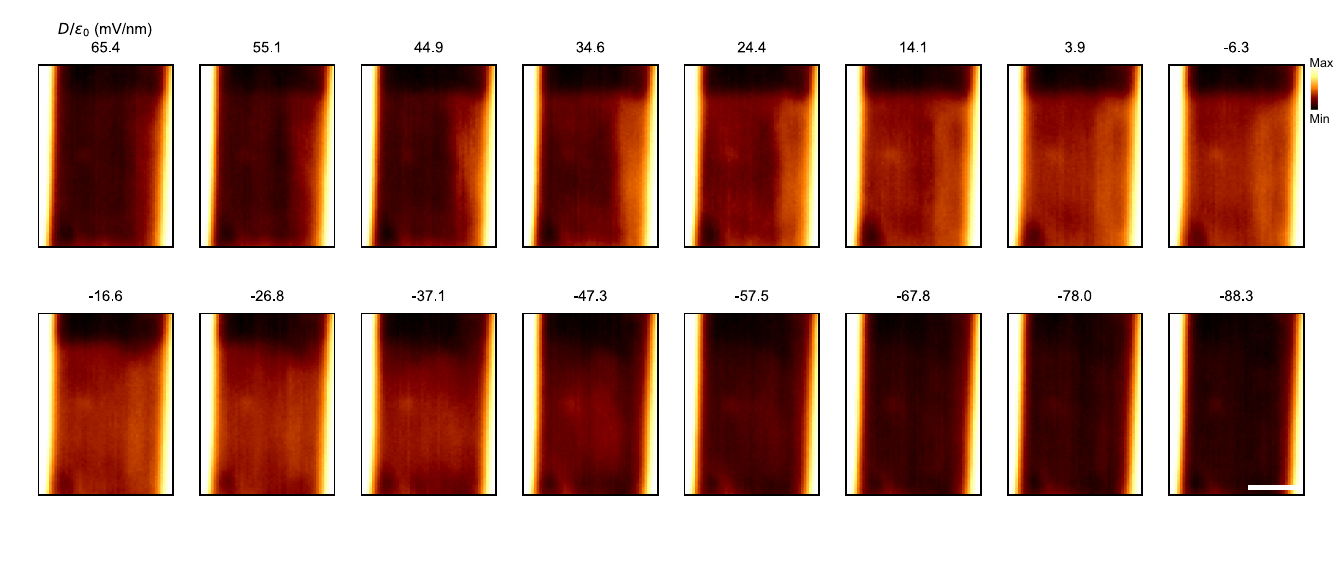}
\caption{\textbf{Real space maps of MIM-Im response at different displacement fields at $\bm{\nu = -1/2}$ filling.}  Real-space maps of the MIM-Im at different displacement fields at $\nu = -1/2$ filling factor measured at $T$ = 1.5~K. Scale bar: 1$\mu$m.  }
 \label{Extended Data fig:6}
 \end{figure}



\clearpage
\renewcommand\thefigure{S\arabic{figure}}
\setcounter{figure}{0}


\end{document}